\documentclass [11pt,a4paper] {article}

\usepackage[cp1252]{inputenc}
\usepackage[english]{babel}

\usepackage{amssymb}
\usepackage{amsmath}
\usepackage{amsfonts,amssymb}

\usepackage[dvips]{graphicx}

\begin{document}

\title{Any realistic theory must be computationally realistic:\\a response to N. Gisin's definition of a Realistic Physics Theory}

\author{Arkady Bolotin\footnote{$Email: arkadyv@bgu.ac.il$} \\ \textit{Ben-Gurion University of the Negev, Beersheba (Israel)}}

\maketitle

\begin{abstract}\noindent It is argued that the recent definition of a realistic physics theory by N. Gisin cannot be considered comprehensive unless it is supplemented with requirement that any realistic theory must be computationally realistic as well.\\
\end{abstract}

\noindent 

\noindent 

\noindent Recently Nicolas Gisin \cite{Gisin} has proposed a definition of realistic physics theories that ``avoids the mere reduction to determinism and avoids the risk of tautology''. According to Gisin's explanation of the meaning of the word ``realistic'', a theory is realistic ``if and only if, according to the mathematical structure of this theory, the collection of all physical quantities written in the system unambiguously determines the probabilities of all possible measurement outcomes.''\\

\noindent Apparently, the main idea behind this definition was to combine already known in the physics literature characterizations of realistic theories in such a way that both classical and quantum theories (including those which predict quantum nonlocality) would be regarded as realistic ones.\\

\noindent Attractive and credible as this definition might appear at first sight, it nevertheless misses one requirement that, in my opinion, each realistic theory must meet. Specifically, Gisin's definition fails to mention that, before anything else, a realistic theory must be \textit{computationally realistic}. This means, that the mathematical structure of a realistic physics theory must allow the collection of all the system's physical quantities to determine the probabilities of all possible measurement outcomes not only in an unambiguous way but\textit{ }also\textit{ in a realistic, reasonable amount of time}.\\

\noindent Undeniably, the defining characteristic of any physics theory is that it makes falsifiable or testable predictions. This assumes that if the model identifying with a particular physics theory were insoluble in a reasonable amount of time (even with access to a supercomputer), then the theory would not have a \textit{realistically} \textit{testable} predictive content, and therefore the term "realistic" would be hardly applicable to this theory.\\

\noindent Of course, the requirement of being computationally realistic would be completely redundant for the definition of a realistic physics theory if every physics model were capable of being solved in reasonable time as soon as enough computational resources were thrown at the model. However, in all likelihood, such is not the actual state of things in our real physical world.\\

\noindent Indeed, as it was shown in the paper \cite{Bolotin}, solving the Schr\"odinger equation for any given Schr\"odinger Hamiltonian is a problem at least as hard as the hardest problems in the \textbf{NP} computational complexity class, which in turn implies that this problem has no efficient (i.e. running in reasonable, not excessive time) algorithm unless the \textbf{P} computational complexity class (which is the set of decision problems solvable on a deterministic computing device within reasonable time) is equal to \textbf{NP}. Since the majority of complexity theorists believes that \textbf{P} $\neq$ \textbf{NP} (see, for instance, \cite{Gasarch}), this means that to solve a quantum-mechanical model of an \textit{arbitrary} system, whose evolution is governed by the Schr\"odinger equation $i\hbar \frac{\partial }{\partial {\rm t}}\left.\left|\psi \left(t\right)\right.\right\rangle =H\left(t\right)\left.\left|\psi \left(t\right)\right.\right\rangle $ (where $H\left(t\right)$ is the Hamiltonian of the system accounting for the kinetic and potential energy of all the particles constituting this system, and $\left.\left|\psi \left(t\right)\right.\right\rangle $ describes the quantum state of the system) is an \textit{intractable problem}. In other words, if the \textbf{P} $\neq$ \textbf{NP} conjecture is true, the computational complexity of the quantum-mechanical model of a system will grow so rapidly with the model's input size (approximately equivalent to the system's constituent particle number) that bringing any additional computational resources to bear on the model will be of no value.\\

\noindent From here, we can infer that among various limitations that might be contained in a possible definition of a realistic physics theory, the requirement of being computationally realistic is perhaps the most stringent one.\\

\noindent Thus, for example, the quantum model of a microscopic system evolving in isolation should be regarded as \textit{a realistic model} since the \textbf{NP}-hard problem of finding the solution to the Schr\"odinger equation for a system that has just a small number of degrees of freedom can be surely solved in a reasonable amount of time. In contrast to this, the quantum model of a truly macroscopic object (such as a macroscopic detector, Schr\"odinger's cat, the universe), whose microscopic degrees of freedom due to the interaction between the internal microscopic particles even with the most coarse grain discretization will be of the same magnitude as a double exponential $2^{N_{\!{\rm A}}}$ of the Avogadro's number $N_{\!{\rm A}}\sim 10^{24}$, ought to be considered as \textit{a non-realistic model}. Obviously, to everyone living in the world of limited in time and space computational resources, the intractable problem possessing that enormous amount of degrees of freedom would be the same as a mere unsolvable problem.\\

\noindent On the other hand, \textbf{NP}-hardness of the Schr\"odinger equation only means that (assuming \textbf{P} $\neq$ \textbf{NP}) this equation does not have an efficient algorithm that solves it \textit{exactly} on each input (for each given Schr\"odinger Hamiltonian). This indicates that if a truly macroscopic object is considered as a system with only a few controlled or measured degrees of freedom (among many others that are uncontrolled or unmeasured) then the object's quantum model (an inexact, probabilistic one as it can only provide an incomplete description of the system) can be surely solved in reasonable time and, as a consequence, regarded as a realistic physics model.\\

\noindent With that being said, I totally agree with Nicolas Gisin that theories, which are non-realistic (particularly, \textit{computationally non-realistic}),  may nevertheless ``exist mathematically and one may wonder whether any future fundamental physics theory may be non-realistic''.\\

\end{document}